\documentclass[journal]{IEEEtran}

\usepackage[cmex10]{amsmath}
\usepackage{array}
\usepackage{mdwmath}
\usepackage{mdwtab}
\usepackage{graphicx}
\usepackage{bbm}
\usepackage{cite}
\usepackage{color}

\newcommand{\vb}[1]{\mathbf{#1}}
\newcommand{\mc}[1]{\mathcal{#1}}
\newcommand{\bmc}[1]{\boldsymbol{\mathcal{#1}}}
\newcommand\sups[1]{^{\hbox{\scriptsize{#1}}}}

\newcommand\subt[1]{_{\hbox{\tiny{#1}}}}
\newcommand{\primedsum}{\sideset{}{'}{\sum}}

\begin{document}

\title{Fluctuation-Induced Phenomena in Nanoscale Systems: Harnessing the Power of Noise}

\author{M.~T.~Homer~Reid\thanks{M. T. Homer Reid is with the Research 
                                Laboratory of Electronics,
                                Massachusetts Institute of Technology.},
        Alejandro~W.~Rodriguez\thanks{Alejandro W. Rodriguez is with the 
                                      School of Engineering and Applied 
                                      Sciences, Harvard University, and the 
                                      Department of Mathematics, Massachusetts 
                                      Institute of Technology.},
        and Steven~G.~Johnson\thanks{S. G. Johnson is with the 
                                     Department of Mathematics,
                                     Massachusetts Institute of Technology.}}

\IEEEspecialpapernotice{(Invited Paper)}

\maketitle

\begin{abstract}
The famous Johnson-Nyquist formula relating noise 
current to conductance has a microscopic generalization 
relating noise current \textit{density} to microscopic 
\textit{conductivity}, with corollary relations
governing noise in the components of the 
electromagnetic fields.
These relations, known collectively in physics as
\textit{fluctuation--dissipation} relations, form
the basis of the modern understanding of
\textit{fluctuation-induced phenomena}, 
a field of burgeoning importance in 
experimental physics and nanotechnology.
In this review, we survey recent progress in computational 
techniques for modeling fluctuation-induced phenomena, 
focusing on two cases of particular interest:
\textit{near-field radiative heat transfer} 
and \textit{Casimir forces}.
In each case we review the basic physics of 
the phenomenon, discuss semi-analytical and 
numerical algorithms for theoretical analysis,
and present recent predictions for novel 
phenomena in complex material and geometric
configurations.
\end{abstract}


%

\section{Introduction}
\IEEEPARstart{E}{very} electrical engineer knows the famous 
Johnson-Nyquist formula for the noise current through a 
resistor,
\begin{equation}
  \big\langle I^2 \big\rangle = 4kT G \Delta f 
\label{IICorrelator}
\end{equation}
where $\langle I^2\rangle$ is the mean-square noise 
current (Fig. 1\textbf{a}),
$kT$ is the temperature in energy 
units, $G=1/R$ is the conductance of the resistor,
and $\Delta f$ is the measurement bandwidth.
Equation (\ref{IICorrelator})---which allows designers
to quantify, and thus compensate for, the unavoidable presence
of noise in physical circuits---is a crucial tool in the 
circuit designer's kit and a mainstay of the electrical 
engineering curriculum from its earliest 
stages~\cite{GrayMeyer2001}.

Perhaps less well-known in the EE community is that 
equation (\ref{IICorrelator}) is only one manifestation of 
a profound and far-reaching principle of 
physics---the \textit{fluctuation-dissipation theorem}---that
relates the mean-square values of various fluctuating
quantities to certain physical parameters (known as 
\textit{generalized susceptibilities}) associated with 
the underlying system. 
In equation (\ref{IICorrelator}),
the fluctuating quantity is the noise current through the 
resistor, and the generalized susceptibility is the 
conductance; more generally, as we will see below,
the fluctuation--dissipation concept allows us to quantify
fluctuations not only in macroscopic device currents but 
also in microscopic current 
\textit{densities},
from which it is a short step to obtain fluctuations in 
the components of the electric and magnetic 
\textit{fields} inside and outside material bodies 
(Fig. 1\textbf{b}). 
In this case, we will see that the key tools turn out to be
nothing but the familiar 
\textit{dyadic Green's functions}, which 
describe the electromagnetic fields of prescribed
current sources 
and are computable by any number of
standard methods of classical electromagnetism.
It is remarkable that many problems in the field of 
fluctuation-induced phenomena, which would at first blush
seem to necessitate complex statistical-mechanical and 
quantum-mechanical reasoning, in fact reduce in practice 
to applications of classical electromagnetic theory 
that would be familiar to any electrical engineer.
\begin{figure}[b]
\centering
\resizebox{21pc}{!}{\includegraphics{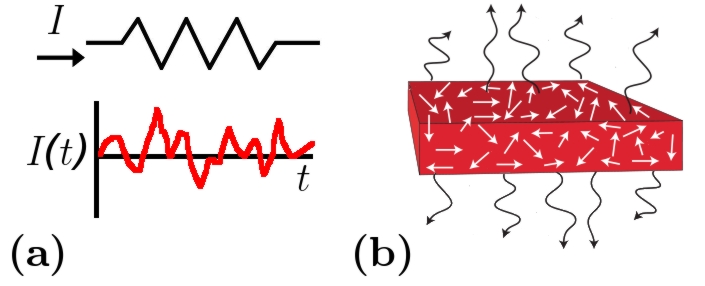}}
\caption{From macroscopic to microscopic noise.
\textbf{(a)} The current through a 
resistor exhibits thermal noise with mean-square amplitude 
proportional to the conductance 
[the Johnson-Nyquist formula, equation (\ref{IICorrelator})].
\textbf{(b)} More generally, the \textit{microscopic}
current density inside a slab of conducting material
exhibits fluctuations with mean-square amplitude
proportional to the microscopic conductivity 
[the fluctuation-dissipation theorem, equation 
(\ref{JJCorrelator})]. Knowledge of these microscopic 
current fluctuations, together with the dyadic Green's
functions of the system, allow us
to predict the mean-square fluctuations in the 
components of the electromagnetic fields in space
[equations (\ref{EHCorrelator}) and (\ref{EECorrelator})].
}
\label{Figure1}
\end{figure}

But why would we \textit{want} to quantify noise in the 
individual components of the electromagnetic fields
around material bodies? The answer is that these
microscopic field fluctuations can mediate 
\textit{macroscopic} transfers of energy or momentum
among the bodies, which become especially dramatic for
bodies at submicron separations.
In the former phenomenon---\textit{near-field radiative heat 
transfer}---fluctuating fields in micron-scale gaps 
between inequal-temperature bodies can lead to a rate of
heat transfer between the bodies that can drastically exceed 
the rate observed at larger separations~\cite{Volokitin2007}.
In the latter phenomenon---the \textit{Casimir effect}---fluctuating
fields around bodies give rise to attractive and repulsive 
forces between the bodies, which generalize the 
familiar van der Waals interactions 
between molecules~\cite{Parsegian2006}.
Both phenomena become negligibly small for bodies separated
by distances of more than a few microns, which places
their observation squarely within the domain of 
nanoscale physics and engineering.

\nocite{Casimir1948} 
\nocite{Boyer1968} 
\nocite{Tomas2002} 
\nocite{Zhou1995} 
\nocite{Emig2004} 
\nocite{Emig2007} 
\nocite{Reid2009} 
\nocite{Levin2010} 
\nocite{Rytov1959} 
\nocite{PolderVanHove1971} 
\nocite{Narayanaswamy2008B} 
\nocite{Rodriguez2011B} 
\nocite{McCauley2011} 

\begin{figure*}
\resizebox{43pc}{!}{\includegraphics{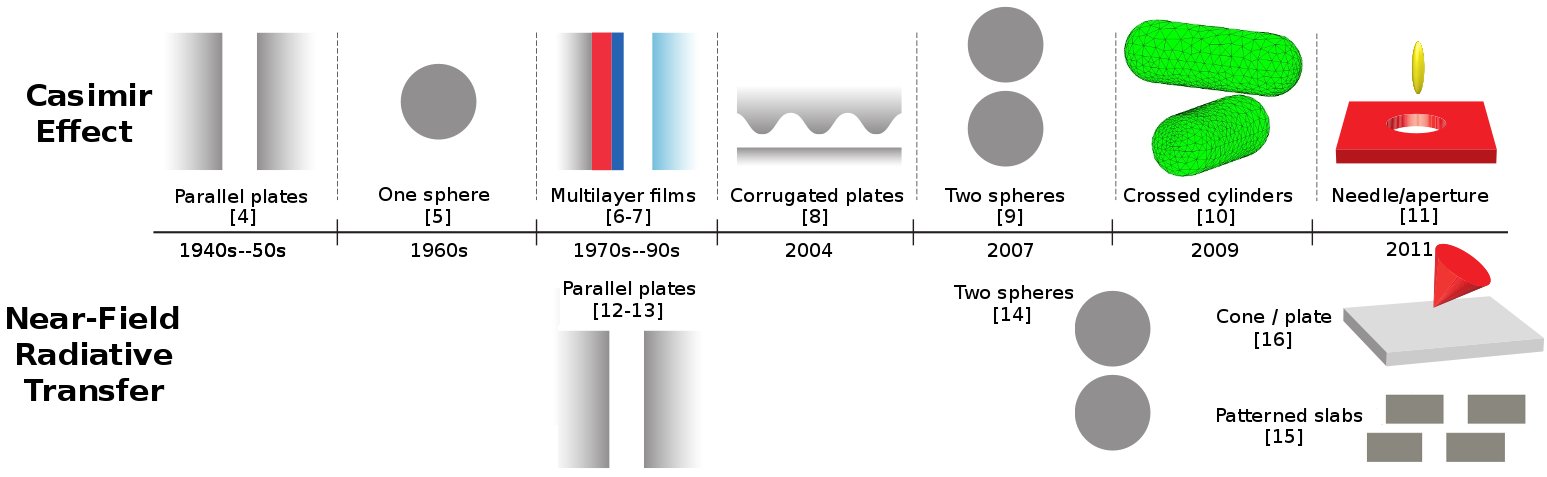}}
\caption{A selective timeline indicating the most complex geometries
for which rigorous calculations of Casimir interactions (upper) or 
near-field radiative heat transfer (lower) were possible at various 
historical epochs. Note that computational techniques such as 
finite-difference grids and boundary-element discretization,
which have been used in electrical engineering for decades, have
only been introduced to the study of fluctuation-induced phenomena
within the past five years.}
\label{TimelineFigure}
\end{figure*}

Although the study of electromagnetic field fluctuations has 
been an active area of physics for decades, its relevance 
to electrical engineering was limited for most
of that time to equation (\ref{IICorrelator}) and other
relations quantifying noise in circuits.
In the past fifteen years, however, this situation has begun
to change; advances in fabrication and measurement technology 
have ushered in a golden age of experimental studies of 
fluctuation-induced 
phenomena~\cite{Volokitin2007,Rodriguez2011A}, and there is 
reason to believe that this fledgling field of experimental physics 
will soon become relevant to electrical engineering in areas such 
as thermal lithography and MEMS technology. This experimental 
progress has created a demand for modeling and simulation tools 
capable of predicting fluctuation phenomena in realistic
experimental configurations, including the complex, asymmetric 
geometries and imperfect materials present in real-world systems.

The evolution of theoretical tools for modeling fluctuation-induced
phenomena mirrors the historical development of techniques for 
solving classical electromagnetic scattering problems.
In the latter case, the earliest calculations were restricted
to highly symmetric geometries (such as Mie's 1908 treatment of 
scattering from spheres) for which a convenient choice of coordinates 
and special-function solutions of Maxwell's equations allow the problem 
to be solved analytically (or at least  
\textit{semi-analytically}---that is, with results obtained as
expansions in special functions, which in practice are then
evaluated numerically~\cite{Harrington1961}).
Later, fully numerical techniques
capable of handling more general geometries 
gradually became available, 
including the finite-difference, finite-element, and
boundary-element methods introduced in the 1960s,
and today the problem of electromagnetic
scattering is addressed by a wealth of comprehensive off-the-shelf
CAD tools capable of handling extremely complex material 
and geometric
configurations.

Advances in the modeling of near-field radiative transfer and
Casimir phenomena have proceeded in similar order
(Fig. \ref{TimelineFigure}). In both cases the first 
calculations were restricted to the simplest parallel-plate 
geometries~\cite{Casimir1948, PolderVanHove1971, Rytov1959};
these were later extended to other simple shapes
such as cylinders~\cite{Emig2004} and  
spheres~\cite{Boyer1968, Klimchitskaya2000, Emig2007, Narayanaswamy2008B, Lambrecht2008, Milton2008A},
and, more recently, 
tools for general geometries have become
available~\cite{Lambrecht2006,Rahi2009,Johnson2010}.
%
%
%
All of these developments, however, have lagged their antecedents
in the classical-scattering domain by many decades; indeed, even
for the relatively simple case of two interacting spheres, the 
Casimir force was only calculated in 2007~\cite{Emig2007} 
and the near-field radiative transfer only in 
2008~\cite{Narayanaswamy2008B}. One reason for this lag 
is the relative paucity of experimental data, which---as 
noted above---are significantly more difficult to gather
for fluctuation-induced phenomena than for classical 
scattering.
But perhaps the main reason that practical computations
of fluctuation-induced phenomena have been so long in coming
is simply that the problems present extraordinary computational
challenges. Indeed, as we will see below, calculations of
near-field radiative-transfer and Casimir phenomena may 
be reduced in practice to the solution of classical scattering
problems---but a great \textit{number}, thousands or even millions,
of separate scattering problems must be solved to compute 
the heat transfer or Casimir force for a single geometric
configuration.

As a result, algorithms for predicting fluctuation phenomena
tend to start with techniques familiar to electrical
engineers---including the $T$-matrix, finite-difference, 
and boundary-element methods of computational 
electromagnetism---but then proceed to combine and modify 
these techniques in novel ways to obtain computational 
procedures that can run in a reasonable length of time.
The goal of this review is to describe these computational
techniques---and some of the results that they have
predicted---in ways that will make sense to electrical
engineers.

\bigskip
\section{Fluctuations in Electromagnetic Sources and Fields: 
         The Johnson--Nyquist Law and Beyond}

The microscopic generalization of equation (\ref{IICorrelator})
is~\cite{CallenWelton1951, PolderVanHove1971}
\begin{align}
&
  \Big\langle J_i(\omega, \vb x) J^*_j(\omega, \vb x^\prime) \Big\rangle
\nonumber \\
&\qquad
  = \frac{1}{\pi}\delta_{ij} \delta(\vb x - \vb x^\prime)
    \left[ \frac{\hbar \omega}{2}
          \coth\left(\frac{\hbar\omega}{2kT}\right)
    \right]
    \sigma(\omega, \vb x)
\label{JJCorrelator}
\end{align}
where $J_i(\omega,\vb x)$ is the $i$th cartesian component of the 
microscopic current density at position $\vb x$ and 
frequency $\omega,$ $\hbar$ is Planck's constant,
and $\sigma(\omega, \vb x)$ is the 
position- and frequency-dependent conductivity.
[$\sigma$ is related to the imaginary part of the 
dielectric permittivity according to~
${\sigma(\omega, \vb x)=\omega\cdot\text{Im }\epsilon(\omega,\vb x)}$; here and
throughout we assume that $\epsilon$ is 
linear and isotropic.] 
\textcolor{black}{In signal-processing language familiar
to electrical engineers, the right-hand side of 
equation (\ref{JJCorrelator}) is the power spectral
density (PSD) of a \textit{colored-noise} process; 
the fact that the PSD is frequency-dependent
(i.e. the fact that this is ``colored'' instead of
white noise) corresponds, in the time domain, to the 
nonvanishing of correlations between currents at 
nearby time points.}

The similarity between equations 
(\ref{IICorrelator}) and (\ref{JJCorrelator}) is obvious:
on the left-hand side we have a mean product of currents,
while on the right-hand side we have a temperature-dependent 
factor and a measure of conductivity. However, the microscopic
equation (\ref{JJCorrelator}) extends the macroscopic
equation (\ref{IICorrelator}) in two important ways.

First, whereas equation (\ref{IICorrelator}) is a 
low-frequency, high-temperature approximation that
neglects quantum-mechanical effects, equation
(\ref{JJCorrelator}) remains valid at all
temperatures and frequencies and explicitly 
\textit{includes} quantum-mechanical effects.
Indeed, taking the low-temperature limit of 
the bracketed factor in (\ref{JJCorrelator}), 
we find 
\begin{subequations}
\begin{equation}
 \lim_{T\to 0}
   \left[
   \frac{\hbar \omega}{2} 
   \coth\left(\frac{\hbar\omega}{2kT}\right)
   \right]
   = \frac{\hbar\omega}{2}
\end{equation}
and equation (\ref{JJCorrelator}) thus predicts
non-zero current fluctuations even at zero
temperature: the well-known quantum-mechanical 
\textit{zero-point fluctuations}. In the high-temperature 
limit, on the other hand, we have  
\begin{equation}
   \lim_{T\to \infty}
   \left[
   \frac{\hbar \omega}{2} 
   \coth\left(\frac{\hbar\omega}{2kT}\right)
   \right]
   = kT;
\end{equation}
\label{EnergyPerMode}
\end{subequations}
this is the \textit{classical} regime, in which  
all dependence on $\hbar$ is lost and we recover the simple 
linear temperature dependence of (\ref{IICorrelator}).
The classical regime is defined by the condition
\begin{equation}
   T \gg \frac{\hbar\omega}{2k} 
   \qquad\text{or}\qquad
   T\text{ in kelvin} \gg \frac{\omega}{4\cdot 10^{12} \text{ rad/s}}, 
\label{HighTThreshold}
\end{equation}
a requirement that in practice is always satisfied in 
circuit-design problems, but which may be readily violated
for infrared and optical frequencies 
($\omega > 10^{15}$ rad/sec). 

The second way in which equation (\ref{JJCorrelator})
extends the reach of the Johnson-Nyquist result is that,
whereas (\ref{IICorrelator}) describes only macroscopic
currents, (\ref{JJCorrelator}) gives information on the
\textit{microscopic} current density, which in turn can
be used to predict fluctuations in the components
of the electric and magnetic fields.
The relevant tools for this purpose are the 
\textit{dyadic Green's functions} (DGFs),
well-known to electrical engineers from problems
ranging from radar and antenna design to microwave 
device modeling~\cite{Harrington1961}.
To recall the definition of these quantities,
suppose we have a material configuration 
characterized by 
spatially-varying linear permittivity and permeability 
functions
$\{\epsilon(\omega, \vb x), \mu(\omega, \vb x)\}.$
(In most of the problems we consider, $\epsilon$ and $\mu$
will be piecewise constant in space.) Then the 
electric DGF describes the field due to a point source
in the presence of the material configuration:
\begin{align}
&G_{ij}\sups{E}\big(\epsilon, \mu; \omega;
                    \vb x, \vb x^\prime\big)
\nonumber \\
&\quad=\left(\text{\parbox{0.32\textwidth}{
        $i$-component of electric field at $\vb x$
        due to a $j$-directed point current source
        at $\vb x^\prime$
        }}
  \right)
\label{ElectricDGF}
\end{align}
while the magnetic DGF $G\sups{M}$ similarly 
gives the magnetic field of a point current source.
(Here and throughout, all fields and currents are 
understood to have time dependence 
$\sim e^{-i\omega t}$.) In (\ref{ElectricDGF})
we have indicated the dependence of $G$ on 
the spatially-varying material properties
$\epsilon$ and $\mu$; the DGFs for a given 
material configuration can be computed using
standard techniques in computational 
electromagnetism, after which the fields
at arbitrary points in space due to a 
prescribed current distribution may be computed
according to
\begin{subequations}
\begin{align}
E_i(\omega, \vb x) 
 &= \int G_{ij}\sups{E}(\omega; \vb x, \vb x^\prime)
         J_j(\omega, \vb x^\prime) d\vb x^\prime
\\
H_i(\omega, \vb x) 
 &= \int G_{ij}\sups{M}(\omega; \vb x, \vb x^\prime)
         J_j(\omega, \vb x^\prime) d\vb x^\prime.
\end{align}
\label{DGFs}
\end{subequations}
\textcolor{black}{Note that the long-range nature of 
the $G$ dyadics ensures that the fields are nonvanishing
even at points $\vb x$ in empty space, i.e. points
at which there are no currents or materials.}

Armed with equations (\ref{JJCorrelator}) and 
(\ref{DGFs}), we can now make predictions for noise 
in the components of the electromagnetic fields. For 
example, the mean Poynting flux at a point $\vb x$ is a sum 
of terms of the form (with $\omega$ arguments 
to $E$, $G$, and $J$ suppressed)
\begin{align}
&\big\langle  E_i(\vb x) H^*_j(\vb x) \big \rangle 
\nonumber\\
&\,\,= 
\int d\vb x^\prime d\vb x^{\prime\prime} \,
 G\sups{E}_{ik}(\vb x, \vb x^\prime)
 G^{\text{\scriptsize{M}}*}_{j\ell}(\vb x, \vb x^{\prime\prime})
\big\langle J_k(\vb x^\prime) J_\ell(\vb x^{\prime\prime})
\big\rangle
\nonumber\\
\intertext{Inserting (\ref{JJCorrelator}),}
&= \int \, d\vb x^\prime
   G\sups{E}_{ik}(\vb x, \vb x^\prime)
   G^{\text{\scriptsize{M}}*}_{jk}(\vb x, \vb x^\prime)
   \Theta\big[\omega, T(\vb x^\prime)\big] 
   \sigma(\omega, \vb x^\prime)
\label{EHCorrelator}
\end{align}
where $T(\vb x)$ is the local temperature 
and $\Theta\big[\omega, T\big]=\frac{\hbar\omega}{2\pi}\coth\hbar\omega/2kT$
is the statistical factor in equation (\ref{JJCorrelator}).
(Summation over repeated tensor indices is implied here and
throughout.)

The obvious advantage of an equation like (\ref{EHCorrelator}) is 
that it reduces a problem in quantum statistical mechanics
(determination of the electromagnetic field fluctuations at
$\vb x$) to a problem in classical electromagnetic scattering 
(computation of the DGFs $G\sups{E,M}$).
The difficulty of this approach lies in the great \textit{number} 
of scattering problems that must be solved. Indeed, equation 
(\ref{EHCorrelator}) says that, to compute the Poynting flux
at a single point $\vb x$, we need the DGFs connecting
$\vb x$ to \textit{all points} $\vb x^\prime$ at which 
the conductivity is nonvanishing; for a typical problem
involving two dissipative bodies in vacuum, this amounts
to a solving a separate scattering problem for each point
in the volume of each body. Moreover, even after completing 
all of these calculations we have still only computed the 
Poynting flux at a single point $\vb x$; in general we will 
want to integrate this flux over a surface to get the total
power transfer at a given frequency, and subsequently to 
integrate over all frequencies to get the total power transfer. 

Thus the fluctuation-dissipation concept, in the form of 
equations (\ref{JJCorrelator}) or (\ref{EHCorrelator}), 
performs the great \textit{conceptual} service of reducing 
predictions of noise phenomena to problems of classical
electromagnetic scattering, but leaves in its wake the 
\textit{practical} problem of how to solve the formidable 
number of scattering problems that result. This difficulty 
has been addressed in a variety of ways, some of which we 
will review in the following sections.

\bigskip
\section{Near-Field Heat Radiation: Fluctuation-Induced 
         Energy Transfer in Nanoscopic Systems}

Fluctuating currents in finite-temperature bodies give rise to 
radiated fields which carry away energy. If there are other 
bodies (or an embedding environment) present at the same temperature,
then any energy lost by one body to radiation is replenished 
by an equal energy absorbed from the radiation of other bodies.
However, between objects at \textit{different} temperatures 
there is a net transfer of power, whose rate we can calculate
in terms of the temperatures and electromagnetic properties of
the bodies.

Historically, the first step in this direction was the 
\textit{Stefan-Boltzmann law}, a triumph of 19th-century 
physics which held that the power radiated per unit surface 
area of a temperature-$T$ body was simply 
$\eta \sigma\subt{SB} T^4$, where $\sigma\subt{SB}$ is a 
universal constant and $\eta$, the \textit{emissivity},
is a dimensionless number between 
0 and 1 characterizing the electrical properties of the 
body (specifically, its propensity to emit radiation relative 
to that of a perfect emitter or \textit{black body}). 
The Stefan-Boltzmann prediction is based on an approximation 
that simplifies the electromagnetic analysis: it considers only 
\textit{propagating} electromagnetic waves, neglecting the 
\textit{evanescent} portions of the $\vb E$ and $\vb H$ 
fields that exist in the vicinity of object surfaces. This
is a good approximation when computing the power transfer 
between a single body and its environment, or
between two inequal-temperature bodies separated
by large distances. 

However, when inequal-temperature bodies are separated by 
short distances, evanescent fields can contribute 
significantly to the Poynting flux and the rate of power 
transfer may deviate significantly from the Stefan-Boltzmann 
prediction. The length scale below which distances are 
to be considered ``short'' is the \textit{thermal wavelength,}
$$\lambda_T=\frac{\hbar c}{kT} 
  \approx 7.6\, \mu\text{m} 
    \cdot \left(\frac{300\text{ K}}{T}\right),
$$ 
and thus, in practice, observing deviations from the Stefan-Boltzmann 
law requires measuring the heat flux between two bodies maintained
at inequal temperatures and at a surface--surface separation of a
few microns. This formidable experimental challenge has recently
been met by several groups~\cite{Volokitin2007, Narayanaswamy2008A}, and this progress
has spurred the development of new theoretical techniques for 
predicting the heat flux between closely-spaced bodies with
realistic material properties and various shapes, which we now 
describe.

\subsection{Radiative Heat Transfer as a
            Scattering Problem}
\label{RadiativeHeatScatteringSubsection}

Consider two homogeneous bodies $\mathcal{B}_{1,2}$
separated by a short distance
and maintained at separate internal thermal equilibria 
at temperatures $T_{1,2}.$ (We will consider
the bodies to exist in vacuum; the case of a finite-temperature
embedding environment is a straightforward generalization.)
The rate at which energy is absorbed or lost by body 1 is given 
as a surface integral of the mean Poynting flux,
\begin{equation}
P_1(\omega)=\frac{1}{2}\int_{\mathcal{S}_1} 
    \Big< \vb E(\omega, \vb x) \times \vb H^*(\omega,\vb x) \Big>
    \cdot d\vb S,
\label{PnOmega1}
\end{equation}
where $\mathcal{S}_1$ is the surface of body 1
(or, equivalently, a fictitious bounding surface containing
body 1 and no other bodies) and $d\vb S$ is the inward-pointing
surface normal. Applying equation~(\ref{EHCorrelator}) reduces 
the quantity in brackets to integrals over the volumes of the 
bodies (again suppressing $\omega$ arguments to $G$):
\begin{align}
&\hspace{-0.7in} P_1(\omega) 
\label{PnOmega2} \\
=\frac{\varepsilon_{ijk}}{2}\int_{\mathcal{S}_1} 
  \bigg\{
  & \sigma_1(\omega) \Theta\big[\omega, T_1\big] 
   \int_{\mathcal{B}_1} G\sups{E}_{i\ell}(\vb x, \vb x^\prime)
                        G^{\text{\scriptsize{M}}*}_{j\ell}(\vb x, \vb x^\prime)
                        \,d\vb x^\prime
\nonumber \\
  &\hspace{-0.4in} +\sigma_2(\omega) \Theta\big[\omega, T_2\big] 
   \int_{\mathcal{B}_2} G\sups{E}_{i\ell}(\vb x, \vb x^\prime)
                        G^{\text{\scriptsize{M}}*}_{j\ell}(\vb x, \vb x^\prime)
                        \,d\vb x^\prime
  \bigg\} \, dS_k.
\nonumber
\end{align}
where $\sigma_{1,2}$ are the conductivities of the bodies.
[Here we have used the Levi-Civita symbol $\varepsilon_{ijk}$
to write the components of the cross product
as $(\vb A \times \vb B)_k = \varepsilon_{ijk}A_i B_j.$]
\textcolor{black}{Note that equation (\ref{PnOmega2}) includes integrations
over the volumes of \textit{both} bodies, since there 
are fluctuating sources present in both bodies. Intuitively
one might expect that reciprocity arguments could be exploited 
to relate the two terms to one another and hence streamline
the calculation to involve integration over just one body;
this intuition is indeed born out in practice, as 
discussed below~\cite{Rodriguez2011B}.}

Equation (\ref{PnOmega2}) reduces the calculation of
the net energy transfer to or from a body to the 
classical electromagnetic scattering problem of computing
the DGFs for a geometry consisting of our two material bodies 
$\mathcal{B}_{1,2}$. The difficulty, as anticipated above, is that
we must solve a great number of scattering problems; 
in principle, for each surface point $\vb x$ and each 
volume point $\vb x^\prime$ in the combined surface--volume 
integrals in (\ref{PnOmega2}) we must solve a separate
scattering problem (computing the fields at $\vb x$ due
to individual point sources at $\vb x^\prime$). 
This challenge is in fact so formidable that computations
for geometries even as simple as two spheres
have only become available in the past few years, 
using techniques which we now review.

\subsection{Semi-Analytical Approaches to Radiative Transfer}
\label{MatrixHeatTransferSection}

A first strategy for evaluating (\ref{PnOmega2}) is to consider
certain highly symmetric geometries for which a convenient 
choice of coordinates allows the DGFs to be evaluated 
analytically.
For example, the earliest near-field heat-transfer 
calculations~\cite{Rytov1959, PolderVanHove1971}
took the two objects to be semi-infinite planar slabs,
in which case the DGFs are analytically calculable.
More recently, several groups have extended this approach
to other highly symmetric geometries in which special-function
expansions of the DGFs are available~\cite{Krueger2011B, Bimonte09, 
Otey11, Narayanaswamy2008B, Messina11A, Messina11B, Guerout2012}. A
particularly convenient tool here is the ``matrix'' approach to 
electromagnetic scattering, a technique first discussed in these
{\sc proceedings} in 1965~\cite{Waterman1965}. 
\textcolor{black}{
To solve scattering problems in this approach, one begins by 
writing down two sets of functions,
$\{\vb E\sups{in}_n(\vb x)\}$ and
$\{\vb E\sups{out}_n(\vb x)\}$; 
these are solutions of Maxwell's equations, in an appropriate
coordinate system, which respectively describe electromagnetic 
waves propagating \textit{inward} from infinity to our 
scattering geometry and \textit{outward} from the scatterer 
into open space.  (For example, in spherical geometries the 
$\{\vb E_n\}$ will be products of vector spherical harmonics 
and spherical Bessel functions~\cite{Harrington1961}.)
The disturbance in the electromagnetic field due to a 
scattering object is then entirely encapsulated
in the object's \textit{T-matrix}, denoted $\mathbbm{T}$,
whose $m,n$ element gives the amplitude of the 
$m$th outgoing wave for a scatterer illuminated by
the $n$th incoming wave. In other words,
\begin{align*}
 \text{if the incident field is } 
   &\vb E\sups{inc}(\vb x)=\vb E_n\sups{in}(\vb x)
\\
 \text{then the scattered field is } 
   &\vb E\sups{scat}(\vb x)=\sum_{m} \mathbbm{T}_{mn} \vb E_m\sups{out}(\vb x).
\end{align*}
} 

\textcolor{black}{
Because the $T$-matrix for a body encodes all information 
needed to understand its scattering properties, it is
often possible to express the solution to a 
radiative-transfer problem in terms of simple matrix
operations on the $T$-matrices of the 
objects involved.} 
As an illustration of the sort of 
concise expression that can result from this procedure,
the methods of Ref.~\cite{Krueger2011B} lead to
a simple trace formula for the spectral density of heat 
radiation from a single sphere at temperature $T$~\cite{Krueger2011C}:
\begin{equation}
   H(\omega,T)
  = -2\Theta^\prime[\omega, T] \,
     \sum_{n} 
      \bigg\{    \text{Re }\mathbbm{T}_{nn}(\omega) 
             + |\mathbbm{T}_{nn}(\omega)|^2 
      \bigg\}
\label{MatrixHeatTransferFormula}
\end{equation}
where $\mathbbm{T}$ is the $T$-matrix of the sphere,
the sum runs over its diagonal elements, and $\Theta^\prime$ is just
$\Theta$ minus the contribution
of the zero-point energy term.

The obvious advantage of an equation like
(\ref{MatrixHeatTransferFormula}) is that it is 
simple enough to be implemented in a few lines of 
{\sc mathematica} or {\sc matlab} for objects 
whose $T$-matrix is known analytically. The difficulty 
is that there are not many such objects; indeed, the only
lossy scatterers for which the $T$-matrix may be obtained
in closed form are spheres, infinite-length cylinders, 
and semi-infinite slabs.
(Idealizing the materials as lossless metals
extends the list of shapes for which the $T$-matrix is
known analytically~\cite{Maghrebi2011}, but this 
is not useful for radiative-transfer problems because 
lossless materials neither absorb nor radiate energy.) To make 
predictions for shapes outside this narrow catalog we must 
turn instead to numerical methods.

\subsection{Numerical Approaches to Radiative Transfer}
\label{NumericalHeatTransferSection}
\begin{figure}
\centering
\resizebox{21pc}{!}{\includegraphics{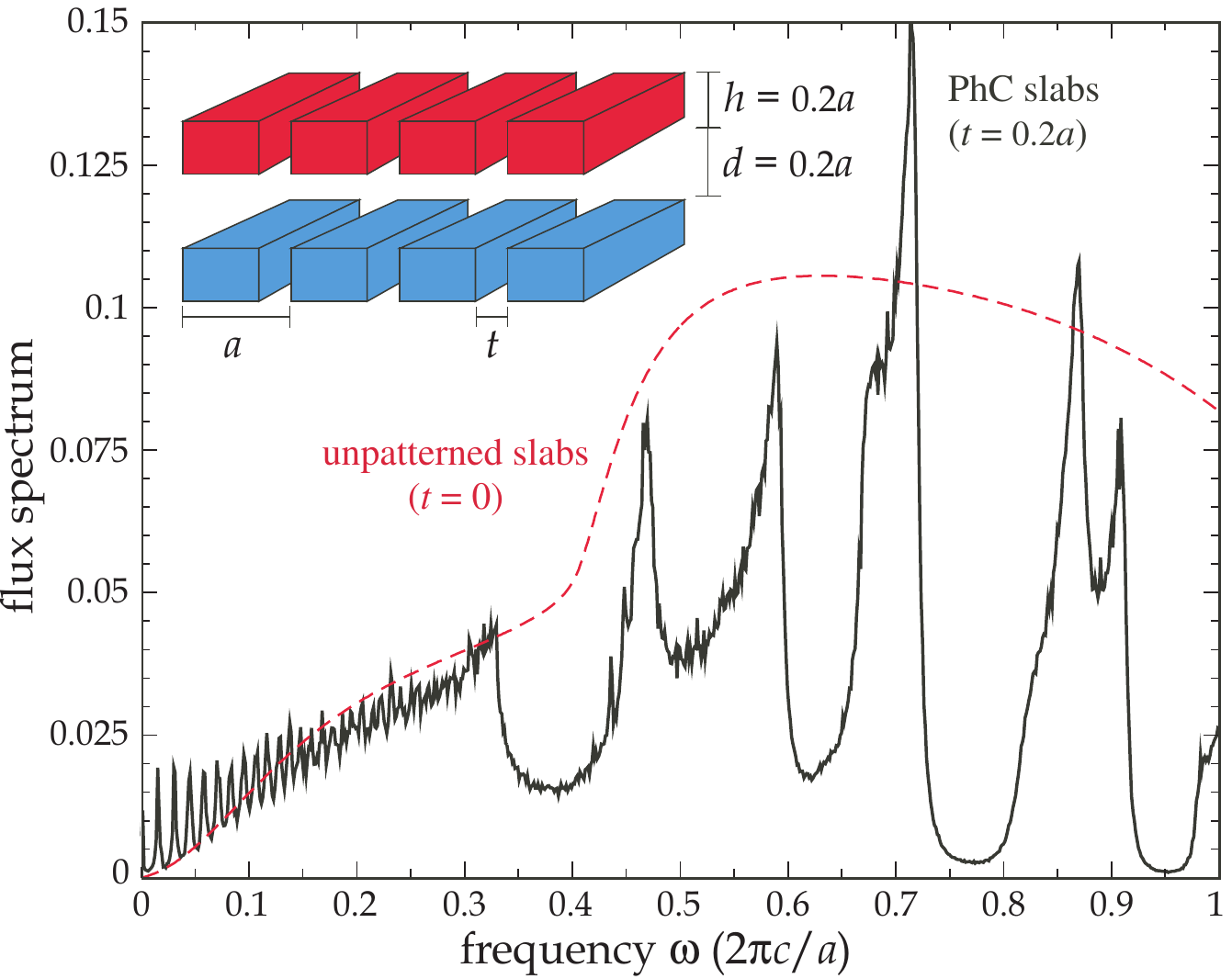}}
\caption{Near-field radiative heat transfer between 
patterned and unpatterned SiC slabs~\cite{Rodriguez2011B}. 
The solid black curve plots the spectral density of power flux 
between SiC photonic crystals (inset) maintained at 
inequal temperatures and surface--surface separation 
$d$. The dashed red curve plots the power flux between
unpatterned SiC slabs.
(In both cases, the power flux is normalized by
the power flux that would obtain between the same
structures at infinite separations $d\to \infty.$)
}
\label{FluxSpectrumFigure}
\end{figure}

One approach to numerical heat-transfer modeling is to combine
matrix-trace formulas in the spirit of equation (\ref{MatrixHeatTransferFormula})
with a numerical technique for computing the $T$-matrices 
of irregularly-shaped objects. This technique was pursued in
Ref.~\cite{McCauley2011}, which investigated heat transfer 
from hot tips of various shapes to a cool planar substrate
at micron-scale distances. In this work, a
boundary-element scattering code was used to compute numerical
$T$-matrices for finite cylinders and finite-length cones; 
a surprising conclusion was that conical tips, 
despite tapering to a point, nonetheless exhibit
\textit{less} spatial concentration (i.e. a larger 
and more diffuse spot size) of heat transferred to the
substrate as compared to cylindrical tips.

An alternative numerical approach to heat-transfer calculations
is to bypass the $T$-matrix approach in favor of a more
direct assault on 
equation (\ref{PnOmega2})~\cite{Luo04, Rodriguez2011B}; here
a ``brute-force'' approach can deliver great generality
with minimal programming time, at the expense of 
much computer time.
Physically, the situation described by equation (\ref{PnOmega2}) 
is that we have randomly fluctuating currents distributed 
throughout the interior of our material bodies, and we wish to 
compute the fields to which these currents give rise.
A particularly convenient way to do this computation
is to run a \text{time-domain} simulation, in which 
we calculate the fields due to a random time-varying 
current distribution whose correlation function in the
frequency domain satifies equation (\ref{JJCorrelator});
by repeating this calculation for many randomly-generated
current distributions and averaging the results, we obtain
approximate ensemble averages of the time-domain
$\vb E$ and $\vb H$ fields, which we may then 
Fourier-analyze to obtain frequency spectra. 
This approach is rendered computationally feasible by exploiting 
several properties of equation (\ref{JJCorrelator}) and of 
Maxwell's equations. 
First, \textit{absence of spatial correlation}: 
the $\delta$ function in (\ref{JJCorrelator}) ensures that 
currents at different locations in space (in particular, currents
in different bodies) are uncorrelated and may thus be chosen 
to have independent random phases. Second, \textit{linearity}: 
although equation (\ref{JJCorrelator}) calls for stochastic 
currents with non-flat spectral density shaped by the factor 
$\Theta[\omega,T]$---what engineers might think of as 
``colored noise''---the linearity of Maxwell's equations ensures 
that we can instead compute the fields due to \textit{white-noise}
currents, which are significantly easier to generate in the
time domain, and only later multiply the resulting frequency 
spectrum by the appropriate shaping factor. Finally, 
\textit{reciprocity}: the flux absorbed by body $\mathcal{B}_2$
due to radiating sources in $\mathcal{B}_1$ is equal to the
flux absorbed by $\mathcal{B}_1$ due to sources in $\mathcal{B}_2$.
This observation allows us to place our stochastic sources only 
in $\mathcal{B}_1$ and compute the resulting flux only into
$\mathcal{B}_2$. 

Combining these arguments leads to a simple expression
for the spectral density of the net heat flux between 
bodies~\cite{Rodriguez2011B}:
$$ H(\omega, T_1, T_2) = \Phi(\omega)
   \Big\{\Theta\big[\omega,T_1\big] - \Theta\big[\omega,T_2\big]\Big\}
$$
where $\Phi$ is the flux into one of the objects due to 
random (white-noise) current sources in the other object.
In practice, $\Phi$ is computed using a finite-difference
time-domain technique, with random current sources placed 
at grid points throughout the volume of the bodies and
the results averaged over many ($\sim$ 60) simulations.

Fig. \ref{FluxSpectrumFigure} illustrates the type of result that
may be obtained using this method~\cite{Rodriguez2011B}. The
solid curve in the figure plots the spectral density of power 
flux between two one-dimensional photonic crystals of SiC 
separated by a short distance $d$ (inset). The dashed curve 
plots the power flux between \textit{unpatterned} SiC slabs. 
(In both cases, the power flux is normalized by the flux between
the same structures at infinite separation $d\to\infty.$)
The patterning of the slabs drastically modifies the flux
spectrum as compared to the unpatterned case.

\bigskip
\section{Casimir Forces: Fluctuation-Induced Momentum Transfer in Nanoscopic Systems}
\label{CasimirEffectSection}

In the previous section, we considered applications of 
fluctuation-dissipation ideas to situations out of 
thermal equilibrium, and we noted the fierce computational 
challenges that arise from the need to solve separate
scattering problems for each point in the volume integration
in (\ref{EHCorrelator}). At thermal equilibrium, a major 
simplification occurs which significantly reduces
computational requirements. The situation is most clearly 
displayed by considering the mean product of $\vb E$-field
components, which reads, in close analogy to 
(\ref{EHCorrelator}),
\begin{align}
&\hspace{-0.1in}
 \big\langle  E_i(\vb x) E^*_j(\vb x^\prime) \big \rangle
\\
&= \int \, d\vb y \,
   G\sups{E}_{ik}(\vb x, \vb y)
   G^{\text{\scriptsize{E}}*}_{jk}(\vb x^\prime, \vb y)
   \Theta\big[\omega, T(\vb y)\big] \sigma(\omega, \vb y)
\nonumber\\
\intertext{The key point is that, at thermal equilibrium, 
$T(\vb y)\equiv T$ is spatially constant, whereupon the 
statistical factor may be pulled out of the integral to yield} 
&= \Theta[\omega, T] \int \, d\vb y\,
   G\sups{E}_{ik}(\vb x, \vb y)
   G^{\text{\scriptsize{E}}*}_{jk}(\vb x^\prime, \vb y)
   \sigma(\omega, \vb y)
\nonumber\\
&= \frac{1}{\omega}\Theta[\omega,T] 
   \, \text{Im }G\sups{E}_{ij}(\vb x, \vb x^\prime).
\label{EECorrelator}
\end{align}
(In going to the last line here we used a standard identity in 
electromagnetic theory which 
follows directly from Maxwell's equations~\cite{Buhmann2009}.)
Thus, evaluating a mean product of field components 
at thermal equilibrium requires the solution of only a 
\textit{single} scattering problem, in contrast to the 
formally infinite number of scattering problems required 
for out-of-equilibrium situations.

Of course, the heat-transfer calculations of the previous section 
are not very interesting at thermal equilibrium, in which by 
definition there can be no net transfer of energy between bodies. 
However, a different sort of fluctuation-induced phenomenon---the
\textit{Casimir effect}---gives rise to nontrivial interactions
among bodies even at the same temperature (and even at zero
temperature), and constitutes a second major branch of the 
study of electromagnetic fluctuations.

\medskip
\subsection{The Casimir Effect}

In 1948~\cite{CasimirPolder1948}, Casimir and Polder
generalized the van der Waals (or ``London dispersion'')
force between fluctuating dipoles of molecules and other
small particles, which depends on the distance $r$ between
the particles like $1/r^7$, to a ``retarded'' force
that varies like $1/r^8$ at large distances (typically
tens of nanometers) where the finite speed of light must
be taken into account.
Later that year~\cite{Casimir1948}, Casimir considered the region 
between two parallel mirrors as a type of electromagnetic cavity,
characterized by a set of cavity-mode frequencies 
$\{\omega(d)\}$ depending on the mirror separation distance
$d$. By summing the zero-point energies 
[equation (\ref{EnergyPerMode}a)] of all modes and 
differentiating with respect to $d$, Casimir predicted an 
attractive pressure between the plates of magnitude 
\begin{equation}
   \frac{F}{A} = \frac{\pi^2 \hbar c}{240 d^4} 
   \approx\frac{10^{-8}\text{ atm}}{(d\text{ in $\mu$m})^4},
\label{CasimirForceMagnitude}
\end{equation}
negligible at macroscopic distances but significant for
surface--surface separations below a few hundred nanometers.

The Casimir effect was subsequently
reinterpreted~\cite{DLP1961, LP1980} as an interaction among
fluctuating charges and currents in material bodies, a 
perspective which allows the use of fluctuation-dissipation
formulas like (\ref{EECorrelator}) to predict Casimir
forces in situations where the cavity-mode picture would
be unwieldy.
In fact, the Casimir effect has been interpreted in a 
bewildering variety of ways; 
in addition to the zero-point-energy picture 
of Ref.~\cite{Casimir1948} and the source-fluctuation 
picture of Ref.~\cite{DLP1961}, 
there are path-integral formulations~\cite{Rahi2009},
world-line methods~\cite{Gies2006}, 
and ray-optics approaches~\cite{Jaffe2004},
to name but a few. Each of these perspectives emphasizes
different aspects of the underlying physics, although
of course all physical interpretations lead ultimately
to mathematically equivalent final results~\cite{Johnson2010}.
However, despite the plethora of theoretical 
perspectives, and even with the simplifications afforded
by thermal equilibrium, the calculations remained so challenging
that force predictions for all but the simplest geometries were 
practically out of reach, and---with experimental progress hampered 
by the difficulty of measuring nanonewton forces between bodies at 
sub-micron distances---for many decades there was little demand for 
computational Casimir methods that could handle general 
geometries and materials.

This situation began to change about 15 years ago with  
the advent of precision Casimir metrology~\cite{Lamoreaux97},
and since that time the Casimir effect has been experimentally
observed in an increasingly wide variety of geometric and 
material configurations (for recent reviews of experimental 
Casimir physics, see~\cite{Capasso07, Rodriguez2011A}).
This experimental progress has spurred the development of
theoretical techniques capable of predicting Casimir forces
and torques in complex, asymmetric geometries with realistic
materials, which we now review.

\subsection{The Casimir Effect as a
            Scattering Problem}

As in Section \ref{RadiativeHeatScatteringSubsection},
we consider two bodies $\mathcal{B}_{1,2}$ in vacuum.
In equation (\ref{PnOmega1}) we integrated the 
average Poynting flux over a surface surrounding a 
body to obtain the rate of energy transfer to that body. 
To compute the rate of \textit{momentum} transfer to
the body---that is, the force on the body---we proceed 
analogously, but now instead of the Poynting flux we integrate
the average Maxwell stress tensor:
\begin{align}
\bmc{F}(\omega)=\int_{\mathcal{S}} 
    \Big< \vb T(\omega, \vb x) \Big> \cdot d\vb S,
\label{FnOmega}
\end{align}
where the components of $\vb T$ are given in terms of 
the components of $\vb E$ and $\vb H$ as 
$$ T_{ij} =
   \epsilon_0 E_i E_j + \mu_0 H_i H_j 
  -\frac{\delta_{ij}}{2}\Big[\epsilon_0 E_k E_k + \mu_0 H_k H_k\Big].
$$
Inserting (\ref{EECorrelator}) and its magnetic analogue
into (\ref{FnOmega}) now yields an expression analogous to 
(\ref{PnOmega2})---but simplified by the absence of volume
integrals---which at temperature $T=0$ takes the form,
for the $i$ component of the force,
\begin{align}
\mathcal{F}_{i}(\omega)=
\frac{\hbar \omega}{\pi}\text{Im}
\int_{\mathcal{S}} 
  \bigg\{ & \epsilon_0\, \mc{G}_{ij}\sups{E}(\omega, \vb x, \vb x) 
          + \mu_0\, \mc{G}_{ij}\sups{H}(\omega, \vb x, \vb x)
\label{FnOmega2}
\\
& \hspace{-0.70in} -\frac{\delta_{ij}}{2} 
     \Big[  \epsilon_0\, \mc{G}_{kk}\sups{E}(\omega, \vb x, \vb x)
           +\mu_0\, \mc{G}_{kk}\sups{H}(\omega, \vb x, \vb x) 
     \Big]
  \bigg\}\,dS_j.
\nonumber
\end{align}
Here $\mc G(\vb x, \vb x^\prime)$,
the \textit{scattering part} of a DGF $G$, is the
contribution to $G$ which remains finite as 
$\vb x^\prime \to \vb x$; this is just the field at 
$\vb x$ due to currents \textit{induced} by a point 
source at $\vb x^\prime$, but neglecting the \textit{direct} 
contribution of that point source. [In (\ref{FnOmega2}), 
$\mc{G}\sups{H}$ is the scattering part of the DGF 
that relates magnetic fields to magnetic currents.]

Equation (\ref{FnOmega2}), like equation (\ref{PnOmega2}),
reduces our problem to that of determining the DGFs for our 
material configuration, and in principle we could now 
proceed to evaluate the surface integral in (\ref{FnOmega2})
with the integrand computed by standard scattering techniques.
For Casimir calculations, however, the situation is
complicated by an important subtlety, which we now discuss.

\subsection{Transition to the Imaginary Frequency Axis}
\begin{figure}
\centering
\resizebox{21pc}{!}{\includegraphics{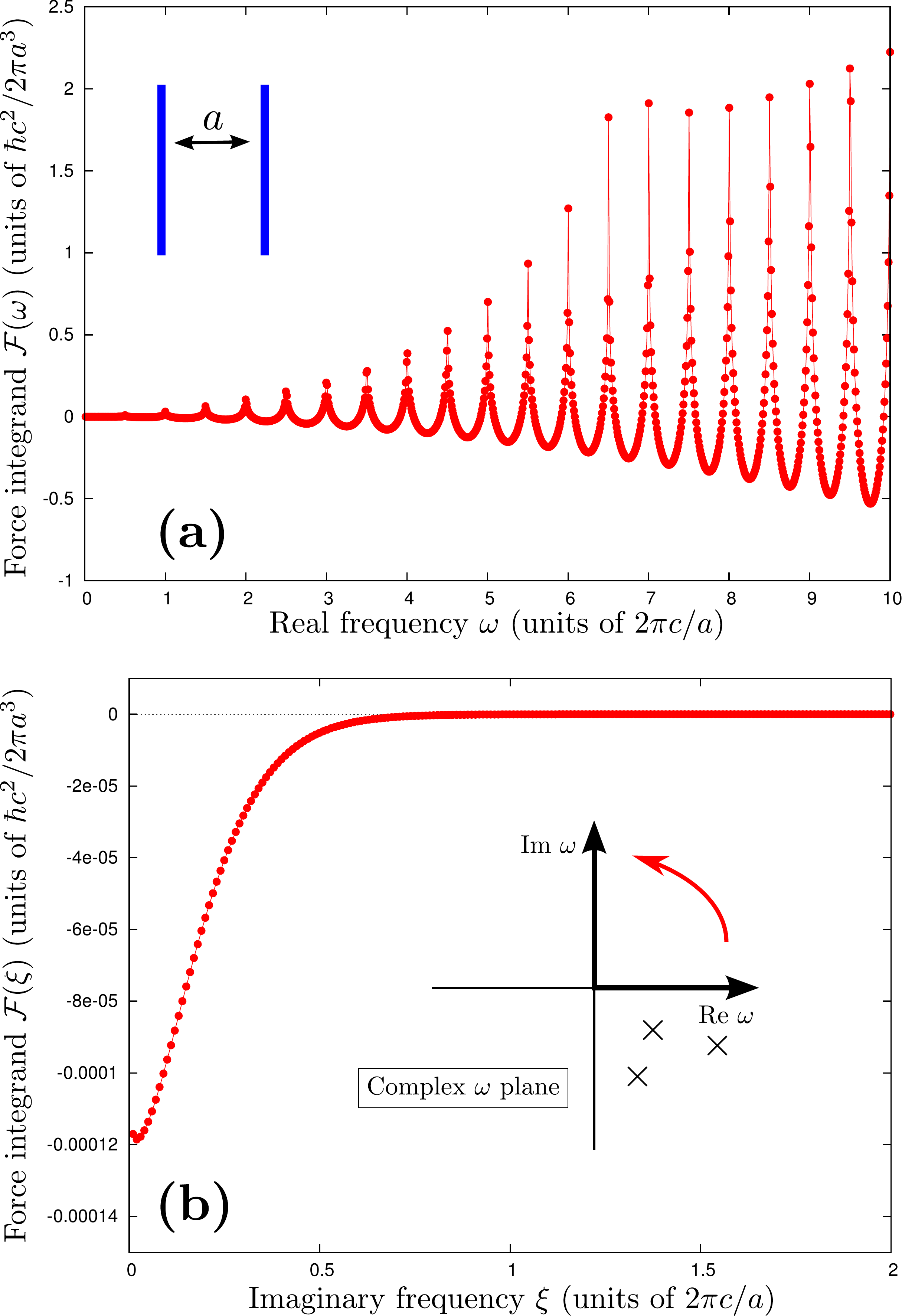}}
\caption{Transition to the imaginary frequency axis.
\textbf{(a)} As a function of real frequency $\omega$, 
the Casimir force integrand $\mathcal{F}(\omega)$ 
of equation (\ref{FnOmega2})---shown here for the 
case of parallel metallic plates separated by a 
distance $a$ (inset)---exhibits severe oscillations
which effectively prohibit evaluation of the
integral (\ref{FIntegral}) by numerical quadrature.
(For the particular case of parallel plates, the
expression for the Casimir force integrand is 
known as the \textit{Lifshitz formula}~\cite{LP1980}.)
\textcolor{black}{
These oscillations are associated with cavity 
resonances, which show up mathematically as 
poles in the lower half of the complex frequency
plane (inset); the real part of the pole corresponds 
to the resonance frequency, while the imaginary part 
corresponds to the width (or the inverse lifetime)
of the resonance.}
\textbf{(b)} Rotating to the \textit{imaginary} 
frequency axis (inset) moves the contour of integration
away from the cavity-resonance poles, resulting in a 
smooth integrand that succumbs readily to numerical 
quadrature.}
\label{ForceIntegrandFigure}
\end{figure}

In contrast to the heat-transfer problems discussed in the 
previous section, for Casimir problems we will not typically
be interested in the contributions of individual frequencies
but will instead seek only the \textit{total} Casimir
force on a body, obtained by integrating (\ref{FnOmega2})
over all frequencies:
\begin{equation}
 F_{i}=
\int_0^{\infty} \mathcal{F}_{i}(\omega)\,d\omega.
\label{FIntegral}
\end{equation}
But na\"ive attempts to evaluate equation (\ref{FIntegral})
numerically are doomed to failure by the existence of rapid
oscillations in the integrand, as pictured in 
Fig. \ref{ForceIntegrandFigure}\textbf{a} for the particular
case of the Casimir force between parallel metallic plates
in vacuum. The origin of these oscillations is not hard to 
identify: they are related to the existence of electromagnetic
\textit{resonances} in our scattering geometry, which 
correspond mathematically to poles of the integrand
in the lower half of the complex $\omega$ plane. (The
oscillatory nature of the force spectrum was emphasized
in Ref.~\cite{Ford1993}, and the implications for 
numerical computations were discussed in 
Ref.~\cite{Rodriguez2007}.)

But this diagnosis of the problem suggests a cure: thinking
of (\ref{FIntegral}) as a contour integral in the complex
frequency plane, we simply rotate the contour of integration 
90 degrees and integrate over the \textit{imaginary} frequency
axis (Fig. 4\textbf{b}). This procedure, known in physics as 
a \textit{Wick rotation}~\cite{Weinberg1996}, yields
\begin{equation}
 F_{i}=
\int_0^{\infty} \mathcal{F}_{i}(\xi)\,d\xi
\label{FIntegral2}
\end{equation}
where $\omega=i\xi$ and $\mathcal{F}$ now involves the DGFs 
evaluated at imaginary frequencies:
\begin{align}
\mathcal{F}_{i}(\xi)=
\frac{\hbar \xi}{\pi}
\int_{\mathcal{S}} 
  \bigg\{ & \epsilon_0 \, \mc{G}_{ij}\sups{E}(\xi, \vb x, \vb x) 
          + \mu_0 \, \mc{G}_{ij}\sups{H}(\xi, \vb x, \vb x)
\label{FnOmega3}
\\
& \hspace{-0.33in} -\frac{\delta_{ij}}{2} 
     \Big[  \epsilon_0 \, \mc{G}_{kk}\sups{E}(\xi, \vb x, \vb x)
           +\mu_0 \, \mc{G}_{kk}\sups{H}(\xi, \vb x, \vb x) 
     \Big]
  \bigg\}\,dS_j.
\nonumber
\end{align}
The Wick rotation is possible here because the DGFs are analytic
functions in the upper half of the complex $\omega$ plane. 
This is a well-known consequence of causality: the fields arise 
after the current fluctuations that generate them~\cite{Jackson1999}.
Another consequence of causality is that, for passive materials,
the permittivity and permeability functions on the
imaginary frequency axis $\{\epsilon(i\xi), \mu(i\xi)\}$
are guaranteed to be real-valued and positive~\cite{Landau1960}.

Physically, the transition to the imaginary frequency axis 
corresponds to replacing the oscillatory time dependence 
$\sim e^{-i\omega t}$ of all fields and currents with an 
exponentially \textit{growing} time dependence $\sim e^{+\xi t}$; 
for frequency-domain computational electromagnetism, this has the 
effect of replacing the spatially oscillatory Helmholtz kernel 
($\frac{e^{i\omega r/c}}{4\pi r}$) with an exponentially 
\textit{decaying} kernel ($\frac{e^{-\xi r/c}}{4\pi r}$).
As illustrated in Fig. 4\textbf{b}, the imaginary-frequency
Casimir force integrand $\mathcal F(\xi)$ is a well-behaved smooth 
function that succumbs readily to numerical quadrature. 

Equations (\ref{FIntegral2}) and (\ref{FnOmega3}) are valid at 
zero temperature. At finite temperatures $T>0$, we must include 
a factor $\Theta[\xi, T]\sim \coth i\hbar\xi/2kT$ under the integral
sign; in this case, it is well-known in physics~\cite{LP1980}
that the integral (\ref{FIntegral2}) over the imaginary frequency 
axis may be evaluated using the method of residues to obtain 
\begin{equation}
F_i = \frac{2\pi kT}{\hbar} \primedsum_{n=0}^\infty \mathcal{F}_i(\xi_n)
\label{FiniteTemperatureForce}
\end{equation}
where $\xi_n = 2n\pi kT/\hbar$, the \textit{Matsubara frequencies},
are just the poles of the $\coth$ factor on the imaginary frequency
axis. [The primed sum in (\ref{FiniteTemperatureForce}) indicates
that the $n=0$ term enters with weight 1/2.]
Computationally, the upshot of equation (\ref{FiniteTemperatureForce}) 
is that finite-temperature Casimir forces are computed with no 
more conceptual difficulty than zero-temperature forces, with the
integral in (\ref{FIntegral2}) simply replaced by the sum in 
(\ref{FiniteTemperatureForce}), although the need to 
evaluate equation (\ref{FnOmega3}) in the limit of zero
frequency ($\xi=0^+$) poses challenges for some methods 
of computational electromagnetism~\cite{Zhao2000, Epstein2010}.
The temperature dependence of Casimir interactions is
a topic of recent theoretical~\cite{Rodriguez2010B}
and experimental~\cite{Shuskov2011} interest.

\subsection{Semi-Analytical Approaches to Casimir Computations}
\label{MatrixCasimirSection}

Like the first studies of near-field radiative transfer,
the first generation of theoretical Casimir 
techniques focused on highly symmetric geometries for which 
analytical scattering solutions are available~\cite{Genet2003, 
Lambrecht2006, Rahi2009, Milton2008B, MaiaNeto2008, 
Davids2010, Bimonte09, Kenneth2008}. 
As an example of the type of concise expression that may be 
obtained via these methods, the zero-temperature Casimir force 
between two compact bodies with center--center separation 
vector $\vb R$ may be expressed in the form~\cite{Emig2007} 
\begin{equation}
F_i
 =
 \frac{\hbar}{2\pi}\int_0^\infty \, \text{Tr }
 \left[
 \mathbbm{M}^{-1}(\xi) \cdot \frac{\partial \mathbbm{M}(\xi)}{\partial \vb R_i}
 \right]
 \, d\xi
\label{EGJKCasimirForce}
\end{equation}
where the matrix $\mathbbm{M}$ has the block structure
$$
   \mathbbm{M}=\left(\begin{array}{cc}
   \mathbbm{T}_1^{-1} & \mathbbm{U}(\vb R) \\
   \mathbbm{U}^\dagger(\vb R) & \mathbbm{T}_2^{-1}
   \end{array}\right);
$$
here $\mathbbm{T}_{n}$ is the $T$-matrix for body $n$ and 
$\mathbbm{U}(\vb R)$ is a \textit{translation} matrix,
which relates spherical Helmholtz solutions about different 
origins and for which closed-form analytic expressions
are available~\cite{Wittmann1988}. [The partial derivative in 
(\ref{EGJKCasimirForce}) is taken with respect to a rigid
displacement of one body in the $i$th cartesian direction.]

Like equation (\ref{MatrixHeatTransferFormula}), 
the formula (\ref{EGJKCasimirForce}) is simple enough 
that it can be implemented in just a few lines of 
{\sc mathematica} or {\sc matlab} code
for geometries in which the $T-$matrix is 
known analytically. Again, however, such geometries are 
rare, and for more complicated geometric configurations
we must turn to numerical methods.

\subsection{Numerical Approaches to Casimir Computations}
\label{NumericalCasimirSection}

The most direct way to apply numerical techniques to Casimir 
computations is simply to evaluate the surface integral in 
(\ref{FnOmega3}) by numerical cubature, with the $\mc{G}$ 
tensors at each integrand point $\vb x$ evaluated by
solving a numerical scattering problem in which we place
a point source at $\vb x$ and compute the scattered
fields back at the same point $\vb x$. In principle, 
this scattering problem may be solved by any of the 
myriad available techniques for numerical solution of 
scattering problems (although the need for imaginary-frequency
calculations poses something of a limitation in practice).
To date, computational Casimir methods based on 
numerical evaluation of (\ref{FnOmega3}) have been
implemented using a variety of standard techniques 
in computational electromagnetism:
the finite-difference frequency-domain 
method~\cite{Rodriguez2007, PasqualiMaggs2008},
the finite-difference time-domain 
method [with some transformations to convert
the integral over frequencies in (\ref{FnOmega3})
into an integral over the time-domain response of a
current pulse]~\cite{Rodriguez2009, McCauley2010, Pan2011},
and the boundary-element 
method~\cite{XiongChew2009, XiongChew2010}. 
\begin{figure}[t]
\centering
\resizebox{21pc}{!}{\includegraphics{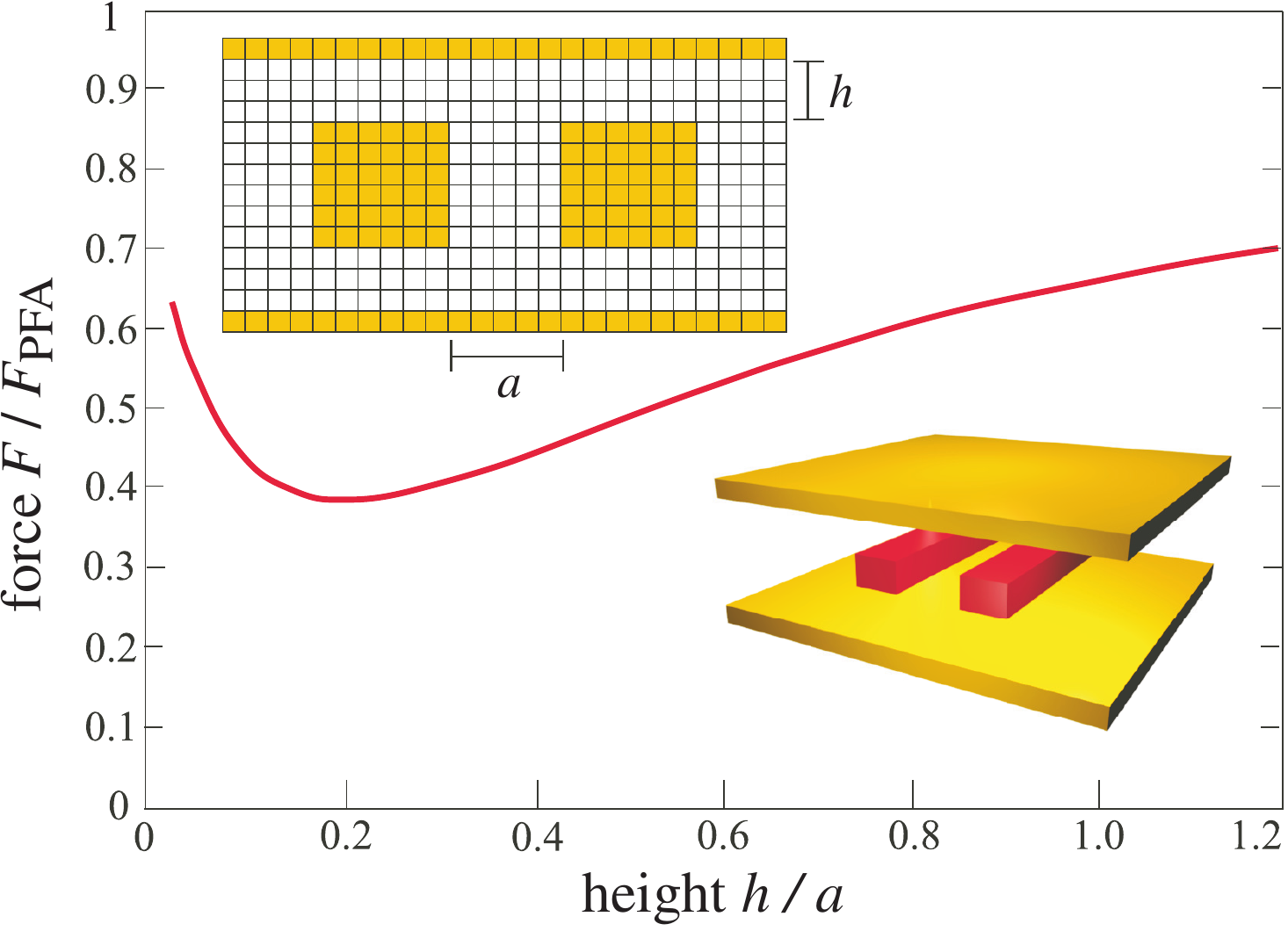}}
\caption{Casimir force between elongated pistons confined between
parallel plates~\cite{Rodriguez2007}.
The lower inset depicts the geometry, while the upper inset 
shows the finite-difference
grid used to model the cross-section of this $z$-invariant
structure. The force between the pistons exhibits a 
suprising \textit{non-monotonic}
dependence on the separation distance $h$ between the pistons
and the plates. [The quantity plotted is the actual force 
divided by the proximity-force approximation (PFA) to the
force, a convenient $h$-independent normalization.]
}
\label{NonMonotonicFigure}
\end{figure}

Compared to the special-function approaches discussed 
in Section \ref{MatrixCasimirSection},
any one of these numerical methods offers the significant
practical advantage of handling arbitrarily complex 
geometries with little more difficulty than simple 
geometries. Among the various numerical methods, the finite-difference
methods have the advantage of greater generality---in the sense
that they can readily handle arbitrarily complex material 
configurations, including anisotropic and continuously-varying 
dielectrics---while the boundary-element methods have the advantage
of greater computational efficiency for the piecewise-homogeneous
material configurations typically encountered in practice.

As an illustration of the type of problem that is facilitated
by numerical Casimir methods, Figure \ref{NonMonotonicFigure} 
plots the force between elongated square pistons confined 
between parallel plates (all bodies are perfect conductors),
as computed using a finite-difference 
technique~\cite{Rodriguez2007}. The lower inset in the figure
depicts the geometry, while the upper inset shows the 
finite-difference grid used to model the cross-section
of this $z$-invariant structure. The force between the pistons 
exhibits a suprising \textit{non-monotonic}
dependence on the separation distance $h$ between the pistons
and the plates.

%
%
\subsection{Fluctuating-Surface-Current Approach to Casimir Computations}
\label{FSCCasimirSection}

The finite-difference and boundary-element methods described 
above have the advantage of great generality,
in that they treat bodies of arbitrarily complex shapes with no 
more difficulty than simple symmetric bodies. However, the need
for numerical evaluation of the surface integral in (\ref{FnOmega3})
adds a layer of conceptual and computational complexity that
is absent from the concise expression (\ref{EGJKCasimirForce}).


An alternative is the recently developed \textit{fluctuating-surface-current} 
approach~\cite{Reid2009, Reid2011A, ReidThesis2011, Reid2012}.
In the FSC technique, we begin with a 
boundary-element-method (BEM) approach to evaluating the DGFs 
in (\ref{FnOmega3}). Instead of proceeding numerically, however,
we exploit the structure of the BEM technique to obtain 
compact analytical expressions for the DGFs in
fully-factorized form, involving products of factors depending
separately on the source and evaluation points.
Inserting these expressions into
(\ref{FnOmega3}) then turns out to allow the surface integral 
to be evaluated \textit{analytically}, in closed form, leaving
behind only straightforward matrix 
manipulations~\cite{ReidThesis2011, Reid2012}. 
The final FSC formula for the Casimir force, 
\begin{equation}
F_i
 =
 \frac{\hbar}{2\pi}\int_0^\infty \text{Tr }
 \left[ 
 \vb{M}^{-1}(\xi) \cdot \frac{\partial \vb {M}(\xi)}{\partial \vb R_i}
 \right]
 \, d\xi,
\label{FSCCasimirForce}
\end{equation}
bears a remarkable similarity to (\ref{EGJKCasimirForce}), but
now with a different matrix $\vb M$ entering into the matrix
manipulations; whereas $\mathbbm{M}$ in (\ref{EGJKCasimirForce})
describes the interactions between incoming and outgoing 
waves in a multipole expansion of the electromagnetic field,
$\vb M$ in (\ref{FSCCasimirForce}) describes the interactions
among \textit{surface currents} flowing on the surfaces of
the interacting objects in a Casimir geometry. [$\vb M(\xi)$
in (\ref{FSCCasimirForce}) is just the usual impedance matrix
that enters into the PMCHW formulation of the boundary-element
method~\cite{Medgyesi1994}, but now evaluated at imaginary frequencies.]
\begin{figure}[t]
\centering
\resizebox{21pc}{!}{\includegraphics{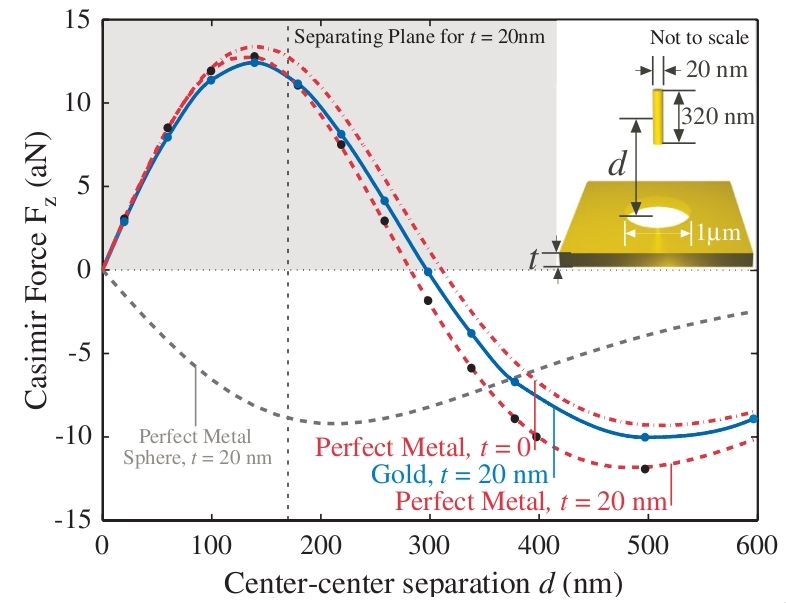}}
\caption{Repulsive Casimir force between metallic objects
in vacuum. Plotted is the $z$-directed force on an 
elongated nanoparticle above a circular aperture in a
metallic plate (inset), as a function of the separation 
distance $d$ between the center of the nanoparticle and 
the center of the plate. The dashed red curves are for the
case of perfectly conducting materials (for two different
plate thicknesses), while the solid blue curve is for the 
case of finite-conductivity gold. The shaded region of the
force curve indicates the repulsive regime, in which 
the nanoparticle is repelled from the plate.
(The dashed vertical line denotes the \textit{separating plane},
i.e. the value of $d$ beyond which the nanoparticle 
is entirely above the plate.)
For comparison, the dashed grey curve indicates the force
on a perfectly-conducting \textit{spherical} nanoparticle; 
in this case the force is attractive at all separations.
(Figure reproduced from Ref.~\cite{Levin2010}.)
}
\label{RepulsiveForceFigure}
\end{figure}

As one example of the type of calculation that is facilitated
by FSC Casimir techniques, Ref.~\cite{Levin2010} investigated 
the Casimir force on an elongated nanoparticle above a circular
aperture in a metallic plate and identified a region of the 
force curve in which the force on the particle is 
\textit{repulsive} (Fig. \ref{RepulsiveForceFigure}). This geometry is notable as the only
known configuration exhibiting repulsive Casimir forces 
between non-interleaved metallic objects in vacuum.
(On the other hand, repulsive forces between dielectric objects
immersed in a dielectric liquid have long been known to exist 
and were observed experimentally in 2009~\cite{Munday2009}; 
in addition, numerical Casimir tools have been used to 
demonstrate theoretically the possibility of achieving stable
suspension of objects in fluids~\cite{Rodriguez2010A},
and further work in this area may have applications
in microfluidics.)

\bigskip
\section{Summary and Outlook}

Despite spending most of its history confined to the realm of pure
physics, the theory and experimental characterization of 
fluctuation-induced electromagnetic phenomena is at last poised
to take on a new role as a growth area in electrical engineering.
The growing ease and ubiquity of nanotechnology are making near-field
radiative transport and Casimir forces increasingly relevant to the
technologies of today and tomorrow, with a corresponding imminent need 
for engineers to account for these phenomena in their designs. In this 
connection it is convenient that a host of powerful 
computational methods, inspired by techniques of classical 
computational electromagnetism but extending these methods
in several ways, have been developed over the past several 
years to model various fluctuation phenomena.
We hope to have convinced the reader that the sudden
conjunction of new theoretical techniques, increasing experimental 
relevance, and the paucity of known results have created burgeoning 
opportunities for computational science---indeed, in fields where two 
spheres represent a novel geometry, the untapped frontiers of
design are vast and inviting.

What lies in store for the future of this field? The work reviewed 
in this article has answered many questions, only to pose 
many more to be addressed in the coming years. Here we give
a brief flavor of some challenges that lie on the horizon.

\textit{General-basis trace formulas for heat transfer.} 
Unlike Casimir forces, the theory of near-field radiation does 
not yet benefit from a compact trace formula that applies to an 
arbitrary localized basis. Existing approaches either require the 
intermediary of a spectral incoming/outgoing wave basis (such as 
cylindrical or spherical waves) that may be ill-suited for 
irregular geometries, or large-scale computations involving costly 
integral evaluations. Is a synthesis (in the spirit of the FSC 
approach of Section \ref{FSCCasimirSection}) 
possible or practical, and what form does it take?

\textit{Fast solvers.} To date, practical applications
of integral-equation Casimir techniques have evaluated the 
matrix operations in equation (\ref{FSCCasimirForce}) 
(matrix inverse, matrix multiplication,
and matrix trace) using methods of 
dense-direct linear algebra. These methods are
appropriate for matrices of moderate dimension 
($D\sim 10^4$ or less), but for larger problems
the $O(D^2)$ memory scaling and $O(D^3)$ CPU-time 
scaling of dense-direct linear algebra renders
calculations intractable.
A similar bottleneck was encountered many years ago
in the computational electromagnetism community,
where it was remedied by the advent of 
\textit{fast solvers}---techniques such as the 
fast multipole~\cite{Greengard1987}
and precorrected-FFT~\cite{Phillips1997} methods that
employ matrix-sparsification techniques to 
reduce the asymptotic complexity scaling 
of matrix operations to more manageable
levels; \textcolor{black}{$O(D^{3/2}\log D$)\cite{Nie2002}
or $O(D\log D)$\cite{Yuan2003,Moselhy2007} are typical}.
Although such
methods could, in principle, be applied to 
stress-tensor Casimir 
computations~\cite{Rodriguez2007,XiongChew2009},
can they be made practical? Can they be applied
to the FSC trace-formula approach, and with what
performance implications?

\textit{New experimental geometries.} Until recently,
theoretical techniques in fluctuation-induced phenomena
lagged behind the forefront of experimental progress
(indeed, as we have seen, it is only in the past few
years that complete theoretical solutions for the simple
sphere--plate geometry commonly seen in experiments have
become available). This situation has recently begun
to change; with a host of new computational methods
for near-field radiative transfer and Casimir phenomena 
becoming available in the past five years, we are 
entering an era in which theoretical predictions can 
be used to guide the design of future 
experiments---and, ultimately, future technologies.
Such a reversal is not without precedent in the history 
of electrical engineering. Indeed, whereas the first 
computational algorithms for modeling antennas and 
transistor circuits were validated by checking that 
they correctly reproduced the behavior of existing 
laboratory systems, today it would be unthinkable to 
fabricate a patch antenna or an integrated 
operational amplifier without
first carefully vetting the design using CAD tools.
Will the development of sophisticated modeling tools 
for near-field radiative transfer and Casimir
phenomena transform those fields as thoroughly as
SPICE and its descendants transformed
circuit engineering? In the former case, 
can we use modeling tools to design efficient
tip--surface geometries for thermal lithography, 
or to invent new solar-cell configurations that 
exploit the interplay of material and geometric 
properties to optimize power absorption and 
retention at solar wavelengths?
In the latter case, can we use computational tools
to understand parasitic Casimir interactions among
moving parts in MEMS devices---or to invent new MEMS
devices that exploit Casimir forces and torques
to useful ends? 

All of these are questions for the future of
fluctuation-induced phenomena. We hope in this 
review to have piqued the curiosity of electrical engineers
in this rapidly developing field---and to have
encouraged readers to stay tuned for future developments.

In closing, we note that all of the computational results 
presented in this review were obtained using 
freely-available open-source software packages for 
computational electromagnetism: 
{\sc meep}, a finite-difference solver, 
and {\sc scuff-em}, a boundary-element 
solver. (Both packages are available for download
at \texttt{http://ab-initio.mit.edu/wiki}.)
In addition to their general applicability to
scattering calculations and other problems in 
computational electromagnetism, these codes offer
specialized modules implementing algorithms discussed 
in this article for numerical modeling of 
fluctuation-induced phenomena.

\section*{Acknowledgments}

This work was supported in part by the Defense Advanced Research
Projects Agency (DARPA) under grant N66001-09-1-2070-DOD, by the Army
Research Office through the Institute for Soldier Nanotechnologies
(ISN) under grant W911NF-07-D-0004, and by the AFOSR Multidisciplinary
Research Program of the University Research Initiative (MURI) for
Complex and Robust On-chip Nanophotonics under grant FA9550-09-1-0704.

\bibliographystyle{IEEEtran}
\bibliography{IEEE201111}

\begin{IEEEbiography}[{\includegraphics[width=1in,height=1.25in,clip,keepaspectratio]{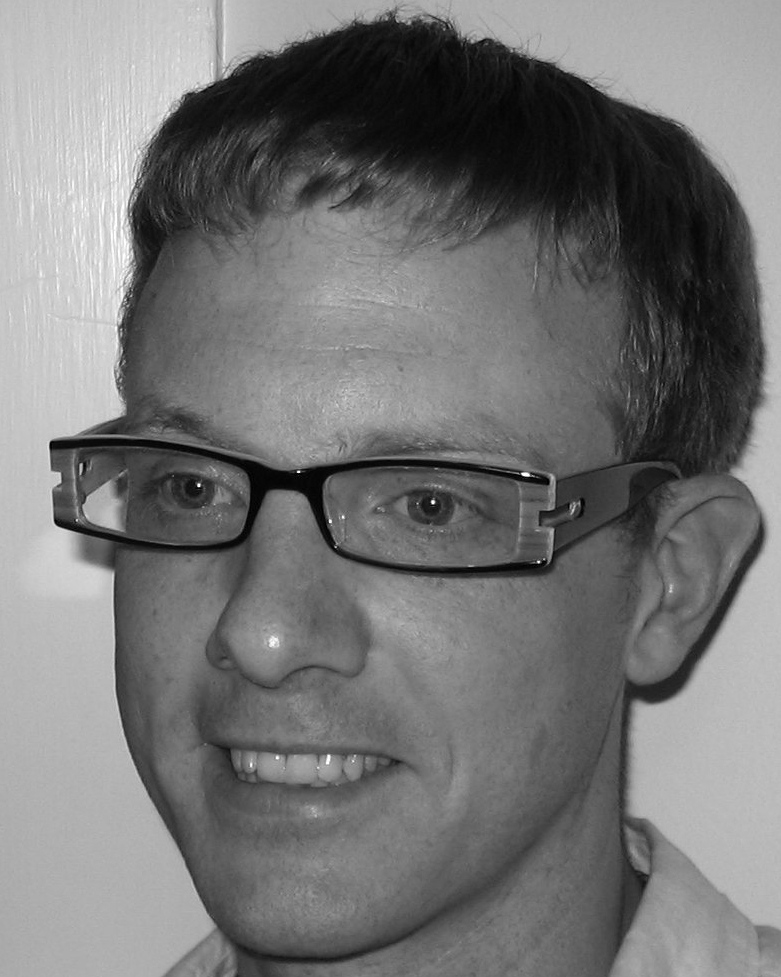}}]{M. T. Homer Reid}
received the B.A. degree in physics from Princeton
University in 1998 and the Ph.D. degree in physics from the Massachusetts
Institute of Technology (MIT) in 2011. From 1998 to 2003 he was Member of
Technical Staff (analog and RF integrated circuit design) at
Lucent Technologies Microelectronics and Agere Systems. He is currently
a postdoctoral research associate in the Research Laboratory of
Electronics at MIT. His research interests include computational 
methods for classical electromagnetism, fluctuation-induced
phenomena, quantum field theory, electronic structure, 
and quantum chemistry. He is 
developer and distributor of {\sc scuff-em}, a free, open-source
software package for boundary-element analysis of problems
in electromagnetism, including nanophotonics, passive RF component
modeling, and Casimir phenomena.
\end{IEEEbiography}

\begin{IEEEbiography}[{\includegraphics[width=1in,height=1.25in,clip,keepaspectratio]{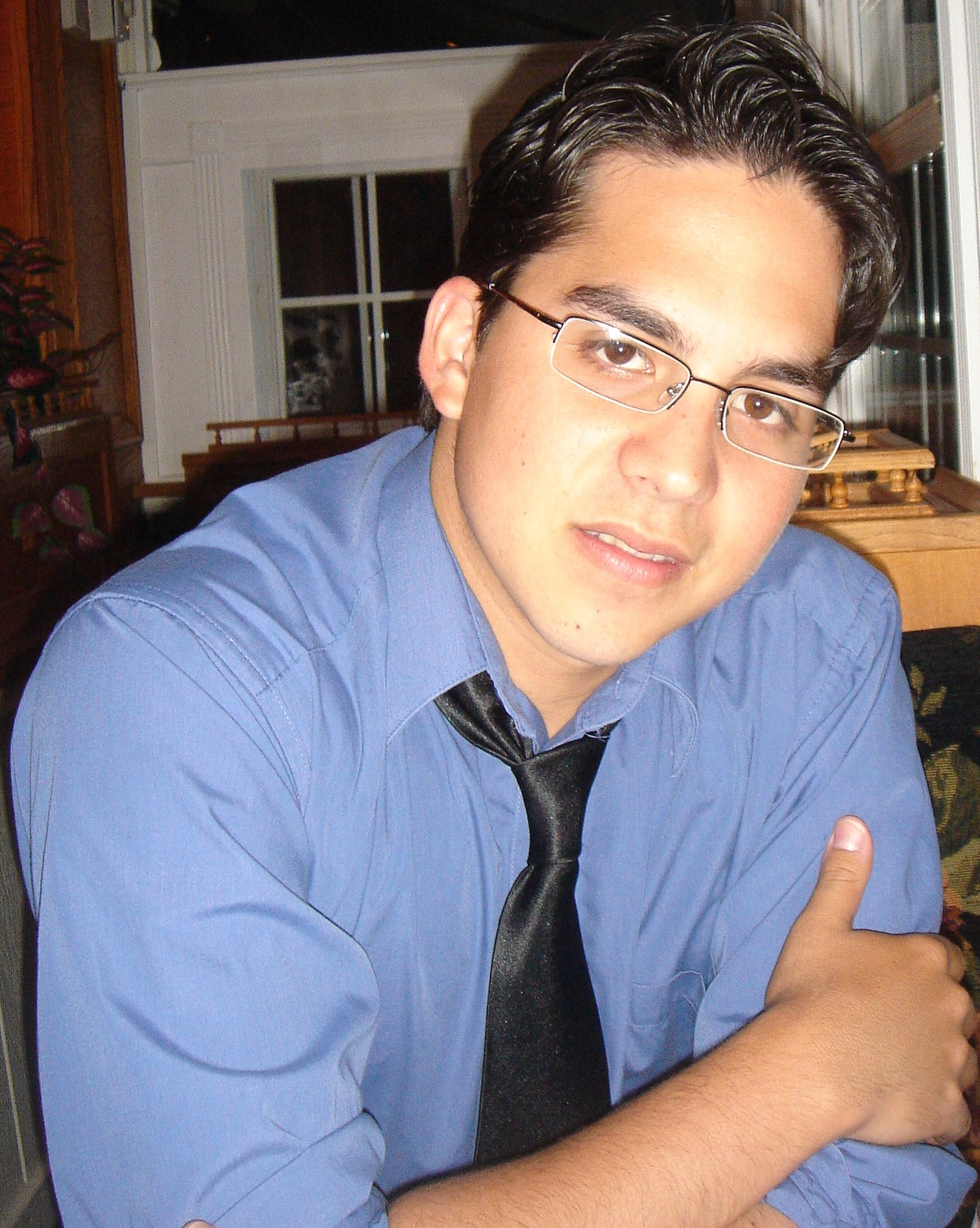}}]{Alejandro W. Rodriguez}
received a B.S. degree in physics from the
Massachusetts Institute of Technology (MIT) in 2006 and a Ph.D. in physics
from MIT in 2010. He is currently a joint Postdoctoral Fellow at the Harvard
School of Engineering and Applied Sciences and at the MIT Department of
Mathematics, working in the areas of computational fluctuation-induced
interactions and nanophotonics. Dr. Rodriguez's work comprise some of the
earliest numerical methods for Casimir calculations, including the first
demonstration of unusual, non-additive, three-body Casimir effects. He is
co-author of over 35 publications and 4 patents in the areas of nonlinear
nanophotonics, Casimir and optomechanical forces, and non-equilibrium
near-field radiative transport. He was named a Department of Energy
Computational Science Graduate Fellow from 2006-2010, was the recipient
of the 2011 Department of Energy Fredrick Howes Award in Computational Science,
and was chosen as a World Economic Forum Global Shaper in 2011. In addition
to his research interests, Dr. Rodriguez participates actively in educational
initiatives aimed at motivating young students to pursue careers in science
and engineering: he was featured in the Spanish-language network Univision,
and on the APS Physics central website, as part of educational campaigns
to increase the number of graduates and underrepresented minorities in STEM
fields, and is currently the professional advisor at the Harvard Society of
Mexican American Engineers and Scientists. A native of Cuba, Dr. Rodriguez
is an avid salsa dancer, film enthusiast, and Cuban hip hop aficionado.
\end{IEEEbiography}

\begin{IEEEbiography}[{\includegraphics[width=1in,height=1.25in,clip,keepaspectratio]{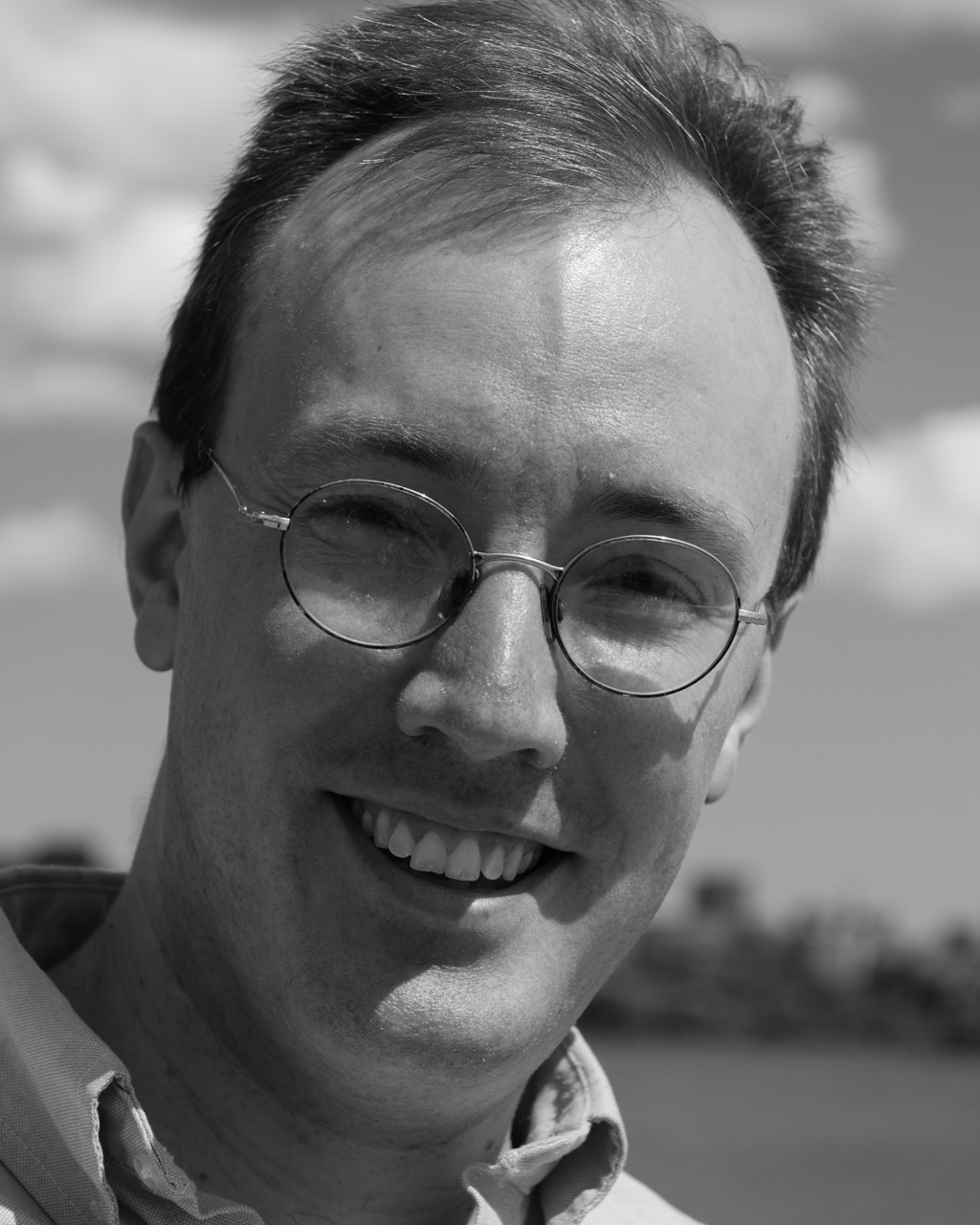}}]{Steven G. Johnson}
received B.S degrees in physics, mathematics, and
computer science from the Massachusetts Institute of Technology (MIT)
in 1995, and a Ph.D. degree in physics from MIT in 2001. Currently, he
is Associate Professor of Applied Mathematics at MIT, where he joined
the faculty of the Department of Mathematics in 2004. He is active in
the field of nanophotonics---electromagnetism in media structured on the
wavelength scale, especially in the infrared and optical regimes---where
he works on many aspects of the theory, design, and computational
modeling of nanophotonic devices. He is co-author of over 150 papers
and over 25 patents, including the second edition of the textbook
Photonic Crystals: Molding the Flow of Light, and was ranked among the
top ten most-cited authors in the field of photonic crystals by
ScienceWatch.com in 2008. Since 2007, his work has extended from
primarily classical nanophotonics into the modeling of interactions
induced by quantum and thermal electromagnetic interactions, and in
particular Casimir forces and near-field radiative transport. In
addition to traditional publications, he distributes several widely
used free-software packages for scientific computation, including the
MPB and Meep electromagnetic simulation tools (cited in over 1000
papers to date) and the FFTW fast Fourier transform library (for which
he received the 1999 J. H. Wilkinson Prize for Numerical Software).
\end{IEEEbiography}

\end{document}